\documentclass[a4paper,12pt]{article}
\usepackage{graphics,cite,amssymb,epsfig,float}
\usepackage[usenames,dvips]{color}

\makeatletter

\@addtoreset{equation}{section}
\makeatother

\def\lsim{\;\raise0.3ex\hbox{$<$\kern-0.75em\raise-1.1ex\hbox{$\sim$}}\;}
\def\gsim{\;\raise0.3ex\hbox{$>$\kern-0.75em\raise-1.1ex\hbox{$\sim$}}\;}

\def\ben{\begin{enumerate}}  \def\een{\end{enumerate}}
\def\bit{\begin{itemize}}    \def\eit{\end{itemize}}
\def\beq{\begin{equation}}   \def\eeq{\end{equation}}
\def\ba{\begin{array}}       \def\ea{\end{array}}
\def\bea{\begin{eqnarray}}   \def\eea{\end{eqnarray}}
\def\nn{\nonumber}

\begin{document}

\begin{titlepage}
\renewcommand{\thefootnote}{\fnsymbol{footnote}}
\setcounter{footnote}{0}

\begin{flushright}
LPT Orsay 08-53 \\
\end{flushright}

\begin{center}
\vspace{3cm}
{\Large\bf Constraints from the Muon g--2 on the Parameter Space}\\ 
\vspace{2mm}
{\Large\bf of the NMSSM} \\
\vspace{2cm}

{\bf Florian Domingo
%\footnote{email: domingo@th.u-psud.fr} 
and 
Ulrich Ellwanger
%\footnote{email: ellwanger@th.u-psud.fr}
} \\
Laboratoire de Physique Th\'eorique\footnote{Unit\'e mixte de Recherche
-- CNRS -- UMR 8627} \\
Universit\'e de Paris XI, F-91405 Orsay Cedex, France
\vspace{2cm}
\end{center}

\begin{abstract}
We generalize the computation of supersymmetric contributions to the
muon anomalous magnetic moment
$(g-2)_{\mu}$ to the NMSSM. In the presence of a light CP-odd Higgs
scalar, these can differ considerably from the MSSM. We discuss the
amount of these contributions in regions of the parameter space of the
general NMSSM compatible with constraints from B physics. In the
mSUGRA-like cNMSSM, constraints from $(g-2)_{\mu}$ prefer regions in
parameter space corresponding to a low SUSY breaking scale.

\end{abstract}

\end{titlepage}

\renewcommand{\thefootnote}{\arabic{footnote}}
\setcounter{footnote}{0}

\section{Introduction}

The result of the measurement of the anomalous magnetic moment of the
muon $a_\mu = (g_\mu-2)/2$ by the E821 experiment at the Brookhaven
National Laboratory (BNL) \cite{bennett} can be considered as a
possible hint for physics beyond the Standard Model (SM).
The determination of the SM contributions to $a_\mu$ requires --
amongst others -- to compute the leading order hadronic contribution
$a_\mu^{HLO}$,  which is the main source of the present SM
uncertainty. Recent  calculations of $a_\mu^{HLO}$ via hadronic $e^+
e^-$ data \cite{jeger1,hagiwara,davier,jeger2,zhang} are in good
agreement, and lead to $\sim 3$~standard deviations between the
experimental value for $a_\mu$ and the SM prediction.

Alternatively, $a_\mu^{HLO}$ can be determined from hadronic
$\tau$-decays \cite{Alemany,Daviertau}, leading to an agreement within
one standard deviation between the experiment and the SM prediction for
$a_\mu$. However, a comparison of the $\tau$ branching fractions into
pions with the corresponding $e^+ e^-$ spectral functions reveals a
discrepancy of $\sim 4.5$ standard deviations \cite{zhang} and requires
assumptions on the pion form factor, isospin violating effects and
vector meson mixings. Since $e^+ e^-$ data is more directly related to
$a_\mu^{HLO}$, it is advocated to use the hadronic $e^+ e^-$ data only
for a computation of $a_\mu^{HLO}$~\cite{jeger2}.

Additional contributions to $a_\mu$ appear within supersymmetric
extensions of the SM (see \cite{stock} for a recent review).
One-loop \cite{1loop,Martin,byrne,martinw} and two-loop
\cite{Degrassi,chang,Cheung,Chen,Arhrib,HSW1,HSW2,fengli,Fengsun}
contributions have been evaluated
within the Minimal Supersym\-metric Extension of the Standard Model
(MSSM). The conclusion is that the MSSM is able to
explain the 3 standard deviations between the experiment and the SM,
provided that the masses of the electroweakly interacting
supersymmetric particles are not far above the electroweak scale
\cite{byrne,martinw, HSW1,HSW2,stock}. On the other hand, the measured
value for $a_\mu$ provides constraints on the parameter space of the
MSSM such as the positivity of the supersymmetric Higgs mass parameter
$\mu$.

The purpose of the present paper is the study of constraints from the
measured value for $a_\mu$ on the parameter space of the
Next-to-Minimal Supersym\-metric Extension of the Standard Model
(NMSSM). First estimates of the contributions to $a_\mu$ in the NMSSM
have been performed in \cite{ghm} (see also \cite{bklh} for $a_\mu$ in a
similar $U(1)'$ model), but subsequently we aim at an
accuracy comparable to the one in the MSSM (limiting ourselves,
however, to dominant logarithms at two-loop order). To this end, most
of the corresponding formulas can be translated in a straightforward
way from the MSSM to the NMSSM.

Nevertheless, the numerical results in the NMSSM can differ
considerably from the ones in the MSSM: the NMSSM contains an
additional gauge singlet superfield $S$, the vacuum expectation value
of which generates the  Higgs mass parameter $\mu$ of the
MSSM~\cite{nmssm}. The CP-even and CP-odd components of $S$ will
generally mix with the neutral components of the two (MSSM) Higgs
doublets $H_u$ and $H_d$. Depending on the parameters of the NMSSM, the
lightest neutral CP-odd Higgs scalar can be quite light \cite{light1}
and lead to numerous new phenomena \cite{light2}. Such a light neutral
Higgs scalar can also have an important impact on $a_\mu$ \cite{kraw}. 
In section 3 we will study possible effects of a light neutral CP-odd
Higgs  scalar on $a_\mu$ in regions of the parameter space of the
NMSSM, which satisfy present bounds from LEP \cite{lep} and $B$-physics
\cite{dom}.

In the remaining part of the introduction we briefly review the various
SM contributions to $a_\mu$ in order to clarify, which additional
contribution would be desirable. The SM contributions to $a_\mu$ can be
split into pure QED contributions $a_\mu^{QED}$ (known to four-loop
order), leading order hadronic contributions $a_\mu^{HLO}$, 
next-to-leading order hadronic contributions $a_\mu^{HNLO}$  (vacuum
polarisation diagrams only) and $a_\mu^{LBL}$ (light-by-light
contributions only), and electroweak effects $a_\mu^{EW}$. The sum is
to be compared with the experimental result \cite{bennett}
\beq\label{1.1e}
a_{\mu}^{EXP}=11\,659\,208.0(5.4)(3.3)\times 10^{-10} \; .
\eeq

The QED contribution is \cite{4lqed}
\beq\label{1.2e}
a_{\mu}^{QED}=11\,658\,471.8113(162) \times 10^{-10}
\eeq
(taking into account estimated five-loop contributions \cite{5lqed})
with an error well below the experimental one. The remaining
difference is
\beq\label{1.3e}
a_{\mu}^{EXP}-a_{\mu}^{QED}=(736.2\pm6.3)\times 10^{-10}\; .
\eeq
Among the recent evaluations of the leading order hadronic
contributions $a_\mu^{HLO}$ \cite{jeger1,hagiwara,davier,jeger2,zhang}
(including the most recent data from SND, CMD-2 and BaBar)
we use -- in order to remain conservative -- the largest estimate from
\cite{jeger1,jeger2}, leading to the smallest deviation from the SM:
\beq\label{1.4e}
a_{\mu}^{HLO} = (692.1\pm 5.6)\times 10^{-10}
\eeq
For $a_{\mu}^{HNLO}$ we use the same reference \cite{jeger1,jeger2}:
\beq\label{1.5e}
a_{\mu}^{HNLO} = (-10.03\pm 0.22)\times 10^{-10}
\eeq
The most recent determination of the hadronic light-by-light
contribution \cite{bijnens} gives
\beq\label{1.6e}
a_{\mu}^{LBL} = (11.0 \pm 4.0) \times 10^{-10} \; .
\eeq

Adding all errors quadratically, one obtains
\beq\label{1.7e}
a_{\mu}^{EXP}-a_{\mu}^{QED+HAD} =(43.1\pm 9.3)\times 10^{-10}\; ;
\eeq
the remaining discrepancy should be explained by electroweak effects
and/or contributions beyond the SM as supersymmetry. Within the SM, the
electroweak contributions are to two-loop order
\cite{czarn}
\beq\label{1.8e}
a_{\mu}^{EW} =(15.4 \pm 0.2)\times 10^{-10}\; ,
\eeq
which leads to the present discrepancy of about three standard
deviations between the experiment and the SM:
\beq\label{1.9e}
a_{\mu}^{EXP}-a_{\mu}^{SM} =(27.7\pm 9.3)\times 10^{-10}\label{deviation}
\eeq

In the next chapter 2 we will review the supersymmetric contributions
to $a_{\mu}$, and specify their dependency on the parameters of the
NMSSM. Contributions to $a_\mu$ which are very similar in the NMSSM
and the MSSM (without possibly light CP-odd Higgs scalars), and
contributions which involve the possibly light pseudoscalar of the
NMSSM, are treated in separate subsections. The formulas of chapter~2
will be made public in the form of a Fortran code on the NMSSMTools web
page \cite{nmssmtools}. In chapter~3 we present numerical results for
$a_{\mu}$ in various regions of the parameter space of the NMSSM,
amongst others regions with a light CP-odd neutral Higgs scalar in the
spectrum, and in the cNMSSM (with universal soft terms at the GUT
scale). At the end of chapter~3 we conclude with a short summary.

\section{\hskip-2mm
Supersymmetric Contributions to $a_{\mu}$ in the NMSSM}

\subsection{Contributions without pseudoscalars}

The one-loop contributions to $a_{\mu}$ in supersymmetric models are
known to consist in char\-gino/\break 
sneutrino or neutralino/smuon loops
\cite{1loop,Martin,byrne,martinw}. The corresponding expressions
are the same in the MSSM and in the NMSSM, provided that the additional
singlino state in the neutralino sector (which includes now five
states) is taken into account. Using the formulas in \cite{Martin}, the
chargino/sneutrino and neutralino/smuon contributions to $a_{\mu}$ can
be written as follows:
\bea
\delta a_{\mu}^{1L\ \chi^\pm} & = & \frac{m_\mu}{16\pi^2}\sum_k
  \left[ \frac{m_\mu}{ 12 m^2_{\tilde\nu_\mu}}
   (|c^L_k|^2+ |c^R_k|^2)F^C_1(x_k)
 +\frac{2m_{\chi^\pm_k}}{3m^2_{\tilde\nu_\mu}}
         {\rm Re}[ c^L_kc^R_k] F^C_2(x_k)\right]\label{amuchar}\\
\delta a_\mu^{1L\ \chi^0} & = & \frac{m_\mu}{16\pi^2}
   \sum_{i,m}\left[ -\frac{m_\mu}{ 12 m^2_{\tilde\mu_m}}
  (|n^L_{im}|^2+ |n^R_{im}|^2)F^N_1(x_{im})
%\right.\nn \\&&\left. \hspace{5cm} 
 +\frac{m_{\chi^0_i}}{3 m^2_{\tilde \mu_m}}
    {\rm Re}[n^L_{im}n^R_{im}] F^N_2(x_{im})\right]\nn\\
    \label{amuneutr} 
\eea

Here $c^R_k$, $c^L_k$, $n^R_{im}$ and $n^L_{im}$ denote the couplings
of the mass eigenstates of the charginos or neutralinos to the
sneutrino or smuons and the muon, which depend on the corresponding
mixing matrices described below:
\bea
c^R_k & = & h_\mu U_{k2} \\
c^L_k & = & -g_2V_{k1}\\
n^R_{im} & = &  \sqrt{2} g_1 N_{i1}
X^{\tilde{\mu}}_{m2} + h_\mu N_{i3}
X^{\tilde{\mu}}_{m1}\\
n^L_{im} & = &  {1\over \sqrt{2}} \left (g_2 N_{i2} + g_1 N_{i1} \right
) X_{m1}^{\tilde{\mu}\,*} - h_\mu N_{i3} X^{\tilde{\mu}\,*}_{m2}
\eea

The conventions for the mass matrices and the resulting mixing matrices
are as in \cite{slha,slha2,Martin,nmssmtools}:
\begin{itemize}
\item The neutralino mass eigenstates $\chi^0_i$, $i=1,\ldots,5$ in the
NMSSM (ordered in increasing absolute mass) are given in terms of the
interaction eigenstates
$\psi_j=$
%\newline 
$(-i\lambda_1,-i\lambda_2,\psi_d^0,\psi_u^0,\psi_s)$
by $\chi_i^0=N_{ij}\psi_j^0$.
\item The chargino mass eigenstates $\chi_k^{\pm}$, $k=1,2$, are
related to the charged gaugino and higgsino  interaction eigenstates
$\psi^+=(-i\lambda^+,\psi_u^+)$ and
$\psi^-=(-i\lambda^-,\psi_d^-)$ through the rotation matrices $U$,
$V$: $\chi^+_k=V_{kl}\psi^+_l$, $\chi^-_k=U_{kl}\psi^-_l$.
\item The smuon mass eigenstates $\tilde{\mu}_m$, $m=1,2$, with masses
$m_{\tilde{\mu}_m}$, result from the interaction eigenstates
$\tilde{\mu}^I=(\tilde{\mu}_L,\tilde{\mu}_R)$ and the rotation
$\tilde{\mu}_m=X^{\tilde{\mu}}_{mn}\tilde{\mu}^I_n$.
In the following, this definition of the matrix $X^{\tilde{\mu}}_{mn}$
will be extended to all sfermions $\tilde{f}$, in which case it will be
denoted as $X^{\tilde{f}}_{mn}$.
\end{itemize}

The $\mu$-sneutrino mass is written as $m_{\tilde\nu_\mu}$. $x_{im}$
and $x_k$ denote the following mass ratios:
$x_{im}=m^2_{\chi^0_i}/m^2_{\tilde\mu_m}$,
$x_k=m^2_{\chi^\pm_k}/m^2_{\tilde\nu_\mu}$. The functions
$F_{1,2}^{C,N}$ are given in the appendix. $h_\mu$ is the muon Yukawa
coupling:  $h_{\mu}=\frac{m_{\mu}}{v_d}$, where $v_d$ is the vacuum
expectation value of the Higgs doublet which couples to down quarks and
leptons.

Since contributions to $a_{\mu}$ require a chirality flip, the formulae
(\ref{amuchar}, \ref{amuneutr}) involve terms -- apart from the
prefactor $m_\mu$ -- which are proportional either to the muon mass
(if the chirality flip occurs in the external legs) or to a
chargino/neutralino mass (when it occurs in internal lines). In
practice, the numerically dominant contributions usually originate from
a chirality flip in the internal lines of the chargino/sneutrino
contribution, i.e. the second term in (\ref{amuchar}). It is
proportional to $c_k^R$, and hence to the Yukawa coupling $h_{\mu}\sim
m_{\mu}/\cos \beta\sim m_{\mu}\tan \beta$ (for $\tan\beta \gg 1$).

An analysis of the chargino/sneutrino diagram reveals (see, e.g.,
\cite{stock}) that this contribution carries the sign of the Higgs
mass parameter $\mu$ ($= \mu_{eff}$ in the NMSSM), hence a positive
value for $\mu$ is phenomenologically favoured. The contribution
decreases with increasing sneutrino mass. For positive $\mu$, the
chargino/sneutrino contribution can well resolve the discrepancy
(\ref{deviation}), provided the muon sneutrino is relatively light or
$\tan \beta$ is large; see, e.g., Figs. 1--3 in section 3. (For a
light sneutrino {\it and} large $\tan\beta$, this contribution can even
be larger than desired.)

It has been pointed out in \cite{Martin,byrne} that the
neutralino/smuon contribution (\ref{amuneutr}) could also explain  the
$3\sigma$ discrepancy (\ref{deviation}), if the bino is quite light and
the smuon mass eigenstates are not too heavy. This bino/smuon 
contribution has the interesting property to be quite insensitive to
$\tan \beta$. 

In the NMSSM, a light neutralino could also be dominantly singlino-like.
However, in this case its contribution to $a_{\mu}$ is suppressed
because of the weak couplings  of the singlino to the MSSM sector.

In addition to these one-loop diagrams, two-loop contributions to
$a_{\mu}$ in the MSSM have been studied in
\cite{Degrassi,chang,Cheung,Chen,Arhrib,HSW1,HSW2,fengli,Fengsun}.
First, we include large logarithms arising from QED corrections to
one-loop diagrams as computed in \cite{Degrassi}: 
\begin{equation}
\delta a_{\mu}^{SUSY+QED}=\delta a_{\mu}^{1L\
SUSY}\left(1-\frac{4\alpha}{\pi}\ln \frac{M_{SUSY}}{m_{\mu}}\right) 
\end{equation} 
This leads to a reduction by a few percents of the LO contributions.

Additional bosonic electroweak two-loop diagrams were computed
in\cite{HSW2}, which can be written as
\begin{equation}
\label{SUSYEW}
\delta a_{\mu}^{2L\,Bos\,EW}=
\frac{5\, G_F\,m_{\mu}^2\,\alpha}{24\sqrt{2}\,\pi^3}
\left(c_L^{2L\,Bos}\ln \frac{m_\mu^2}{M_W^2}+c_0^{2L\,Bos}\right)\; .
\end{equation}
(Up to the more complicated Higgs sector of SUSY models, (\ref{SUSYEW}) 
contains two-loop electroweak SM contributions included in
(\ref{1.8e}). We took care not to count the SM contribution twice.)
Subsequently we confine ourselves to leading logarithmic contribution
$\sim c_L^{2L\,Bos}$, which reads \cite{HSW2}
\begin{equation}\label{cl1}
c_L^{2L\,Bos}=\frac{1}{30}
\left[98+9c_L^h+23\left(1-4s_W^2\right)^2\right]\;.
\end{equation}

In the MSSM, the Higgs contribution $c_L^h$ is of the form (see
eq.~(27) in \cite{HSW2})
\beq\label{cl2}
c_L^h = \frac{\cos{2\beta} M_Z^2}{\cos\beta} \left[ \frac{\cos\alpha
\cos(\alpha+\beta)}{m_H^2} + \frac{\sin\alpha
\sin(\alpha+\beta)}{m_h^2} \right]\; ,
\eeq
where $m_{h,H}$ and $\alpha$ are the CP-even Higgs masses and the
mixing angle in the MSSM. In order to generalize the Higgs contribution
to the NMSSM, it is convenient to re-write eq.~(\ref{cl2}) in terms of
elements of the (inverse) CP-even Higgs mass matrix $M_S^{-2}$, here in
the basis ($H_{u}$, $H_{d}$):
\beq\label{cl3}
c_L^h = \cos{2\beta} M_Z^2 \left[ \left(M_S^{-2}\right)_{22} - 
\tan\beta \left(M_S^{-2}\right)_{12} \right]
\eeq
(Now it is straightforward to verify the sum rule $c_L^h=1$ \cite{HSW2}
with the help of the tree level mass matrix $M_S^2$.)

Eq. (\ref{cl3}) can be interpreted as the result of the evaluation of
the Higgs-dependent two-loop diagrams, where Feynman rules in the
interaction basis ($H_{u}$, $H_{d}$) have been used and mass insertions
were treated perturbatively -- this procedure reproduces the leading
logarithms. Furthermore, in the interaction basis it is easy to
identify the additional contributions from the singlet scalar of the
NMSSM, which decouples from the muon as well as from gauge bosons. The
only possible additional contributions arise from the coupling of the
singlet to charged Higgs bosons (involving always at least one power of
the Higgs singlet-doublet coupling $\lambda$), and the simultaneous
presence of a mass insertion transforming a doublet into a singlet
(also of ${\cal O}(\lambda)$). Since the coefficient of $c_L^h$ in
(\ref{SUSYEW}) is only $\sim -2\times 10^{-11}$, we neglect
subsequently effects of ${\cal O}(\lambda^2/g^2)$ in $c_L^h$. Then eq.
(\ref{cl3}) is equally valid in the NMSSM; all that remains to be done
is to replace $M_S^2$ by the $3\times 3$ mass matrix of the CP-even
sector of the NMSSM. Returning, for convenience, to mass eigenstates,
$c_L^h$ is then of the form
\begin{equation}
c_L^h=\cos 2\beta M_Z^2
\sum_{i=1}^3\frac{S_{i2}\left(S_{i2}-\tan\beta S_{i1}\right)}
{m_{h_i}^2}\; .
\end{equation}
Here, the conventions for the Higgs mixing matrices $S_{ij}$ are as
follows \cite{nmssmtools} (for convenience we discuss the complete
Higgs sector here, which is useful for what follows below; note that
$H_u$ and $H_d$ are exchanged w.r.t. the conventions in
\cite{slha,slha2}):
\begin{itemize}
\item The CP even Higgs eigenstate $h_i$, $i=1,2,3$ (ordered in
increasing mass) in the NMSSM are a mixture of the real parts of the
neutral Higgs components $S^I=(H_{uR},H_{dR},S_R)$: $h_i=S_{ij}S^I_j$.
\item The CP odd Higgs states $a_i$, $i=1,2$ in the NMSSM originate
from the imaginary parts of the neutral Higgs components
$P^I=(H_{uI},H_{dI},S_I)$ (after omission of the Goldstone boson 
$G^0=-\sin \beta H_{uI}+\cos \beta H_{dI}$):
$a_i=P'_{i1}\left(\cos\beta H_{uI}+\sin \beta
H_{d_I}\right)+P'_{i2}S_I$.
\item The charged Higgs boson $H^{\pm}$ is obtained from the charged
Higgs components  $H_u^{\pm}$, $H_d^{\pm}$ as $H^{\pm}=\cos \beta
H_u^{\pm}+\sin \beta H_d^{\pm}$.
\end{itemize}

Two-loop diagrams involving closed sfermion and chargino loops were
studied in \cite{Arhrib,HSW1} and \cite{HSW2}, respectively. The
leading contributions (photonic Barr-Zee diagrams) can be found in
\cite{Arhrib,stock} in the context of the MSSM and can be generalized
to the NMSSM through a replacement of the corresponding couplings.

For the sfermionic diagrams we use the formula (5) of \cite{Arhrib}:
\begin{equation}\label{Sfer}
\delta a_\mu^{2L\,\tilde{f}}=\frac{G_F m_\mu^2 \alpha}{4 \sqrt{2}
\pi^3} \sum_{\tilde{f}}\sum_{i=1}^3
N_c^{\tilde{f}}Q_{\tilde{f}}^2
\frac{\mbox{Re}[\lambda_{\mu}^{h_i}\lambda_{\tilde{f}}^{h_i}]}
{m_{\tilde{f}}^2}f_{\tilde{f}}
\left(\frac{m^2_{\tilde{f}}}{m^2_{h_i}}\right)
\end{equation}
where $N_c^{\tilde{f}}$ corresponds to the number of colors (3 for
squarks, 1 for sleptons), $Q_{\tilde{f}}$ is the electric charge, and
the Higgs/sfermion couplings in the NMSSM are given by:
\begin{eqnarray}\label{Stop}
\lambda_{\tilde{T}_k}^{h_i} &=&
\frac{2\sqrt{2}M_W}{g_2}
\left[h_t\left\{A_tS_{i1}-\lambda\left(sS_{i2}+v_dS_{i3}\right)\right\}
\mbox{Re}[X^{\tilde{T}\,*}_{k1}X^{\tilde{T}}_{k2}]\right.
\nonumber\\ & &\qquad \qquad 
+\left\{h_t^2v_uS_{i1}-\frac{g_1^2}{3}
 \left(v_uS_{i1}-v_dS_{i2}\right)\right\}\left|X^{\tilde{T}}_{k2}
 \right|^2\nonumber\\
 & &\qquad \qquad  \left.+\left\{h_t^2v_uS_{i1}
 -\frac{3g_2^2-g_1^2}{12}\left(v_uS_{i1}-v_dS_{i2}\right)\right\}
 \left|X^{\tilde{T}}_{k1}\right|^2\right]
\end{eqnarray}
\begin{eqnarray}\label{Sbot}
\lambda_{\tilde{B}_k}^{h_i}&=&\frac{2\sqrt{2}M_W}{g_2}
\left[h_b\left\{A_bS_{i2}-\lambda\left(sS_{i1}+v_uS_{i3}\right)\right\}
\mbox{Re}[X^{\tilde{B}\,*}_{k1}X^{\tilde{B}}_{k2}]\right.\nonumber\\
 & &\qquad \qquad  +\left\{h_b^2v_dS_{i2}+\frac{g_1^2}{6}
 \left(v_uS_{i1}-v_dS_{i2}\right)\right\}
 \left|X^{\tilde{B}}_{k2}\right|^2\nonumber\\
 & &\qquad \qquad  \left.+\left\{h_b^2v_dS_{i2}
 +\frac{3g_2^2+g_1^2}{12}\left(v_uS_{i1}-v_dS_{i2}\right)\right\}
 \left|X^{\tilde{B}}_{k1}\right|^2\right]
\end{eqnarray}
\begin{eqnarray}\label{Stau}
\lambda_{\tilde{\tau}_k}^{h_i}&=&\frac{2\sqrt{2}M_W}{g_2}
\left[h_{\tau}\left\{A_{\tau}S_{i2}-\lambda\left(sS_{i1}+v_uS_{i3}
\right)\right\}\mbox{Re}[X^{\tilde{\tau}\,*}_{k1}X^{\tilde{\tau}}_{k2}]
\right.\nonumber\\
 & &\qquad \qquad  +\left\{h_{\tau}^2v_dS_{i2}
 +\frac{g_1^2}{2}\left(v_uS_{i1}-v_dS_{i2}\right)\right\}
 \left|X^{\tilde{\tau}}_{k2}\right|^2\nonumber\\
 & &\qquad \qquad  \left.+\left\{h_{\tau}^2v_dS_{i2}
 +\frac{g_2^2-g_1^2}{4}\left(v_uS_{i1}-v_dS_{i2}\right)\right\}
 \left|X^{\tilde{\tau}}_{k1}\right|^2\right]
\end{eqnarray}
while the muon/Higgs couplings are simply $\lambda_{\mu}^{h_i}
=\frac{S_{i2}}{\cos\beta}$. $v_u$, $v_d$ and $s$ denote the vacuum
expectation values of $H_u$, $H_d$ and $S$, respectively, in the
normalisation where $v_u^2 + v_d^2 = 1/(2\sqrt{2}G_F)$. $\lambda$ is the
Higgs singlet-doublet Yukawa coupling, and $A_{t,b,\tau}$ are the soft
supersymmetry breaking Higgs-sfermion trilinear couplings for the third
generation.

For the closed chargino loop, we employ the formula (63) of
\cite{stock}:
\bea
  \delta a_\mu^{2L\,\chi^{\pm}} &=&
  \frac{G_F m_\mu^2 \alpha}{4 \sqrt{2} \pi^3}
 \sum_{k=1}^2\left[\sum_{i=1}^3\frac{\mbox{Re}
 [\lambda_{\mu}^{h_i}\lambda_{\chi_k^{\pm}}^{h_i}]}
 {m_{\chi_k^{\pm}}}f_{S}
 \left(\frac{m^2_{\chi_k^{\pm}}}{m^2_{h_i}}\right)
%\right. \nn \\&& \qquad \qquad \quad\ \left. 
+ \sum_{i=1}^2 \frac{\mbox{Re}
 [\lambda_{\mu}^{a_i}\lambda_{\chi_k^{\pm}}^{a_i}]}
{m_{\chi_k^{\pm}}}f_{PS}
\left(\frac{m^2_{\chi_k^{\pm}}}{m^2_{a_i}}\right)\right]
\nn\\ &&\label{char}
\eea
with the chargino/Higgs couplings of the NMSSM:
\begin{eqnarray}
\lambda_{\chi^{\pm}_k}^{h_i}&=&\frac{\sqrt{2}M_W}{g_2}
\left[\lambda U_{k2}V_{k2}S_{i3}+g_2
\left(U_{k1}V_{k2}S_{i1}+U_{k2}V_{k1}S_{i2}\right)\right]\\
\lambda_{\chi^{\pm}_k}^{a_i}&=&\frac{\sqrt{2}M_W}{g_2}
\left[\lambda U_{k2}V_{k2}P'_{i2}-g_2
\left(U_{k1}V_{k2}\cos\beta+U_{k2}V_{k1}\sin\beta\right)P'_{i1}\right]
\end{eqnarray}
The functions $f_{\tilde{f}}$, $f_S$ and $f_{PS}$ can be found in the
appendix.

Among the missing contributions are one-loop diagrams with the exchange
of a Higgs boson between the muons, and two-loop diagrams with a closed
SM fermion loop involving Higgs bosons (and a photon). Since these can
play a particular r\^ole in the NMSSM, they will be treated separately
in the next subsection.

\subsection{Contributions including pseudoscalars}
 
Higgs effects are usually negligibly small in the SM or the MSSM
because of the existing lower bounds on the Higgs masses. The SM 
one-loop Higgs/muon diagram is, indeed, about four orders of magnitude
below the sensitivity of the BNL experiment for a SM Higgs mass above
$114$~GeV. However, this is not necessarily the case in the NMSSM:
while the lightest CP-even Higgs boson mass cannot be far below the LEP
bound unless it decouples from the SM sector, the lightest CP-odd boson
$a_1$ can be as light as a few GeV \cite{light1,light2}. Bounds from
B-physics\cite{dom}, especially from $BR(B_s\rightarrow \mu^+\mu^-)$,
can still be satisfied for low values of $\tan\beta$ or when the
loop-induced $b-s-a_1$ coupling is suppressed.

In the context of two-Higgs-doublet-models, the impact of light
(pseudo-) scalars on $a_{\mu}$ was already pointed out in 
\cite{chang, Cheung,kraw}. In the NMSSM, a short analysis of
$a_{\mu}$ has been performed in \cite{ghm}.

In the following we include contributions from the Higgs sector to
$a_{\mu}$ up to the two-loop level. One-loop scalar/fermion diagrams
have been known for quite a while; see, e.g., \cite{Leveille}. For
completeness we detail all the SUSY Higgs contributions (CP-odd, even
and charged):
\begin{eqnarray}
\delta a_{\mu}^{1L\ CP\,even}&=&\frac{G_\mu
m_{\mu}^2}{4\sqrt{2}\pi^2} \sum_{i=1}^3 
\frac{S_{i2}^2}{ \cos^2\beta}\int_0^1{\frac{x^2(2-x)\,dx}{x^2+
\left(\frac{m_{h_i}}{m_{\mu}}\right)^2(1-x)}}\\
\delta a_{\mu}^{1L\ CP\,odd}&=&-\frac{G_\mu
m_{\mu}^2}{4\sqrt{2}\pi^2}
\sum_{i=1}^2 
P'^2_{i1}\tan^2\beta\int_0^1{\frac{x^3\,dx}{x^2+
\left(\frac{m_{a_i}}{m_{\mu}}\right)^2(1-x)}}\label{amuhiggs1L}\\
\delta a_{\mu}^{1L\ charged}&=&
\frac{G_\mu
m_{\mu}^2}{4\sqrt{2}\pi^2}\tan^2\beta 
\int_0^1{\frac{x(x-1)\,dx}{x-1+
\left(\frac{m_{H^{\pm}}}{m_{\mu}}\right)^2}}
\end{eqnarray}

For the two-loop Higgs diagrams involving a closed SM fermion loop (in
contrast to closed sfermion/chargino loops considered in the previous
subsection), we follow the analysis of \cite{Cheung} and generalize it
to the NMSSM:
\begin{eqnarray}
\delta a_{\mu}^{2L\ CP\, even}
&=&\frac{G_{\mu}m_{\mu}^2 \alpha}{4\sqrt{2}\pi^3}
\sum_{i=1}^3
\left[\frac{4}{3}\frac{S_{i1}S_{i2}}{\sin\beta
\cos\beta}f_S\left(\frac{m_t^2}{m_{h_i}^2}\right)\right.\nn\\
 & &\qquad
\left.+\frac{1}{3}\frac{S_{i2}^2}{\cos^2\beta}f_S
\left(\frac{m_b^2}{m_{h_i}^2}\right)+\frac{S_{i2}^2}{\cos^2\beta}f_S
\left(\frac{m_{\tau}^2}{m_{h_i}^2}\right)\right]\\
\delta a_{\mu}^{2L\ CP\,odd}&=&\frac{G_{\mu}m_{\mu}^2 \alpha}
{4\sqrt{2}\pi^3}\sum_{i=1}^2
P'^2_{i1}\left[\frac{4}{3}f_{PS}
\left(\frac{m_t^2}{m_{a_i}^2}\right)\right.\label{amuhiggs2L}\nn\\
& & \left.\qquad +\tan^2\beta\left\{\frac{1}{3}
f_{PS}\left(\frac{m_b^2}{m_{a_i}^2}\right)+
f_{PS}\left(\frac{m_{\tau}^2}{m_{a_i}^2}\right)\right\}\right]
\end{eqnarray}
where the functions $f_S$, $f_{PS}$ are defined in the appendix.

As noticed in \cite{kraw}, one-loop and two-loop light Higgs
contributions are of opposite signs and interfere, therefore,
destructively. In the case of a CP-odd scalar, the one-loop
contribution is negative and worsens the discrepancy (\ref{deviation})
correspondingly. However, for a light CP-odd Higgs heavier than $\sim
3$ GeV, the positive two-loop contribution is numerically more
important. The sum of both contributions is maximal around
$m_{a_1}\sim6$~GeV, though fairly constant in the range $4-10$~GeV (see
Fig.~5 below). Both one- and two-loop contributions are proportional to
the product of two muon Yukawa couplings, which leads to an enhancement
quadratic in $\tan\beta$. They are also proportional to the square of
the mixing of the light pseudoscalar to the doublet sector ($\sim
(P'_{11}\sin \beta)^2$, if we consider the dominant $H_d$ component
only). In spite of the appearance of $P'_{11}$, this coupling can be
large enough to allow the light pseudoscalar contribution to reach the
experimental $2\sigma$ range of $a_\mu$ by itself, provided that
$\tan\beta \gsim 30$. Hence, this contribution can alleviate the upper
bound on slepton masses, which can be derived under the assumption that
the chargino contribution (\ref{amuchar}) explains the
deviation~(\ref{deviation}).

Finally, we estimate the theoretical uncertainty following the analysis
of \cite{stock}, allowing for a 2\% error on the one-loop contributions
and a relative error of 30\% for the  two-loop results. Except for the
light pseudoscalar contributions, which we include in the error
computation described above, we do not expect large sources of
uncertainties different from the MSSM. Therefore, we use the same
additional constant terms as \cite{stock} for the evaluation of the
error.

\section{Results}

In this section we discuss a few phenomenological aspects of the
supersymmetric contributions to $a_{\mu}$ discussed above. First we
consider the $\tan\beta$ and slepton mass dependences in NMSSM
scenarios without a light CP-odd scalar; as expected, these results are
essentially the same as in the MSSM. Then we examine the case of a
light pseudoscalar and show how relevant the Higgs contribution may
become. Finally we study the cNMSSM (with universal soft terms at the
GUT scale), and conclude with a short summary.

\begin{figure}[ht]
\begin{center}
\includegraphics[width=13 cm]{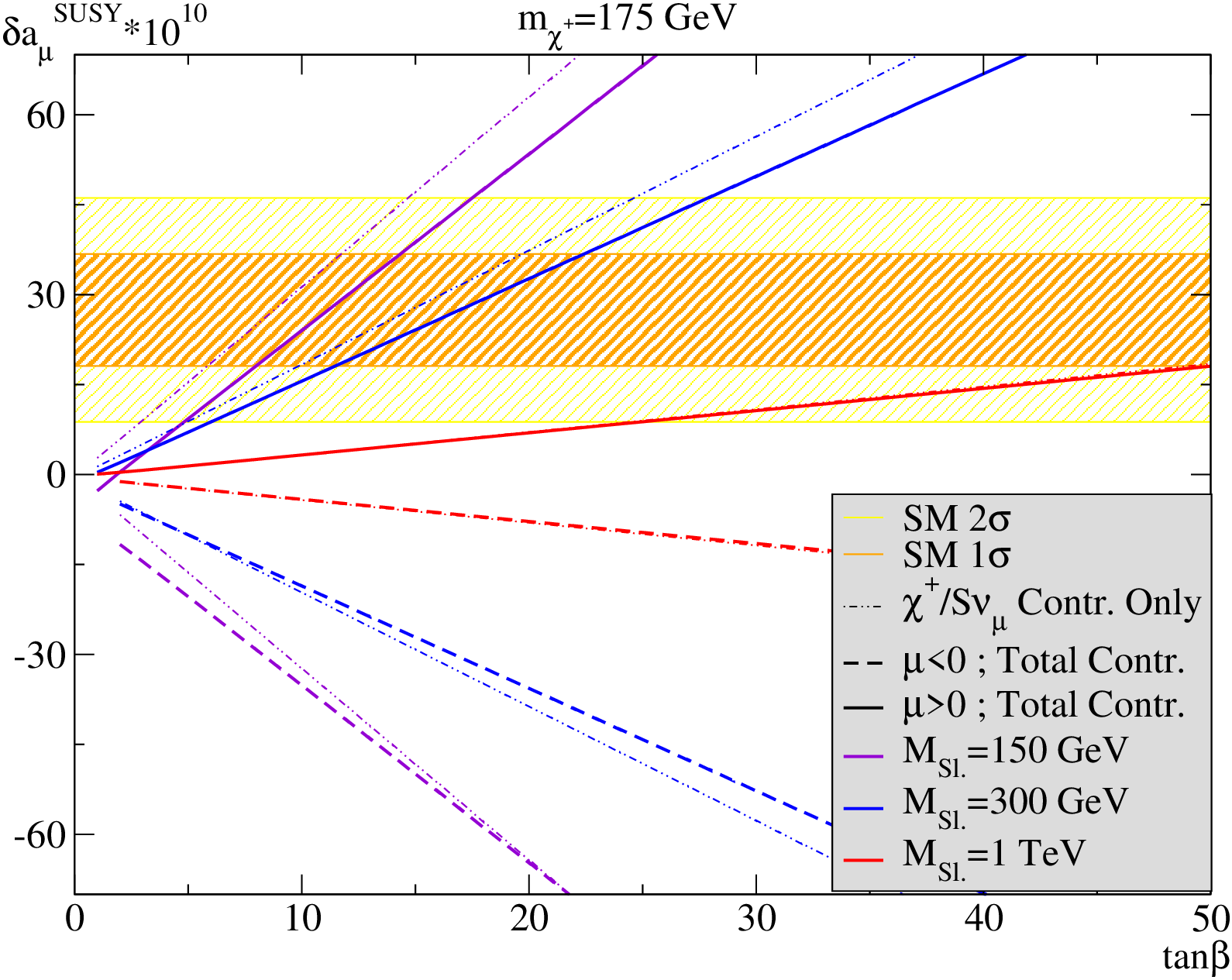}
\end{center}
\caption{The SUSY contribution to $a_{\mu}$ as a function of
$\tan\beta$ for various slepton (sneutrino/smuon) masses.}
\end{figure}

In Fig. 1, we plot the supersymmetric contribution $a_{\mu}^{SUSY}$ as
a function of $\tan\beta$ for various slepton (sneutrino/smuon) masses.
Here we chose, for simplicity, universal soft slepton masses $M_{SL}$
at low energy.
The gaugino soft terms are assumed to be hierarchical
($M_3=3M_2=6M_1=900$~GeV) and $\mu_{eff}$ is chosen such that the
lighter chargino $\chi^+$ has a mass of 175~GeV.
The experimentally allowed  $1\sigma$ and $2\sigma$ regions are
indicated as an orange and a yellow band, respectively. The full
violet, blue or red curves (corresponding to $\mu_{eff} > 0$) and
dashed violet, blue or red curves (corresponding to $\mu_{eff} < 0$)
include all SUSY contributions.  Next to them we plot as dot-dashed
lines the one-loop chargino/sneutrino contribution separately
(corrected by large QED logarithms; for sneutrino/smuon masses of 1~TeV
they differ hardly from the full SUSY contributions). As we mentioned
in the previous section, the chargino/sneutrino contribution is
proportional to the Yukawa coupling of the muon and thus to
$\tan\beta$. Since this contribution obviously dominates the total SUSY
contribution (the small difference is mainly due to the neutralino
diagram), the total SUSY contribution is also roughly proportional to
$\tan\beta$.

The chargino contribution to $a_{\mu}$ carries the same sign as the
SUSY parameter $\mu$ ($\mu_{eff}$ in the NMSSM). Hence, the case
$\mu<0$ (dashed curves in Fig. 1) is disfavoured by the sign of the
difference of the BNL result and the SM (\ref{deviation}). When $\mu$
is positive (full curves in Fig.~1), the SUSY contribution to $a_{\mu}$
is able to account for the $3\sigma$ deviation. When sleptons are heavy
(the red curve), however, large values of $\tan\beta$ are necessary. On
the other hand, for light sleptons (violet curve) the chargino
contribution can be large enough even for moderate values of
$\tan\beta$, or even too important when $\tan\beta$ is large.

\vspace{10mm}
\begin{figure}[ht!]
\begin{center}
\includegraphics[width=13 cm]{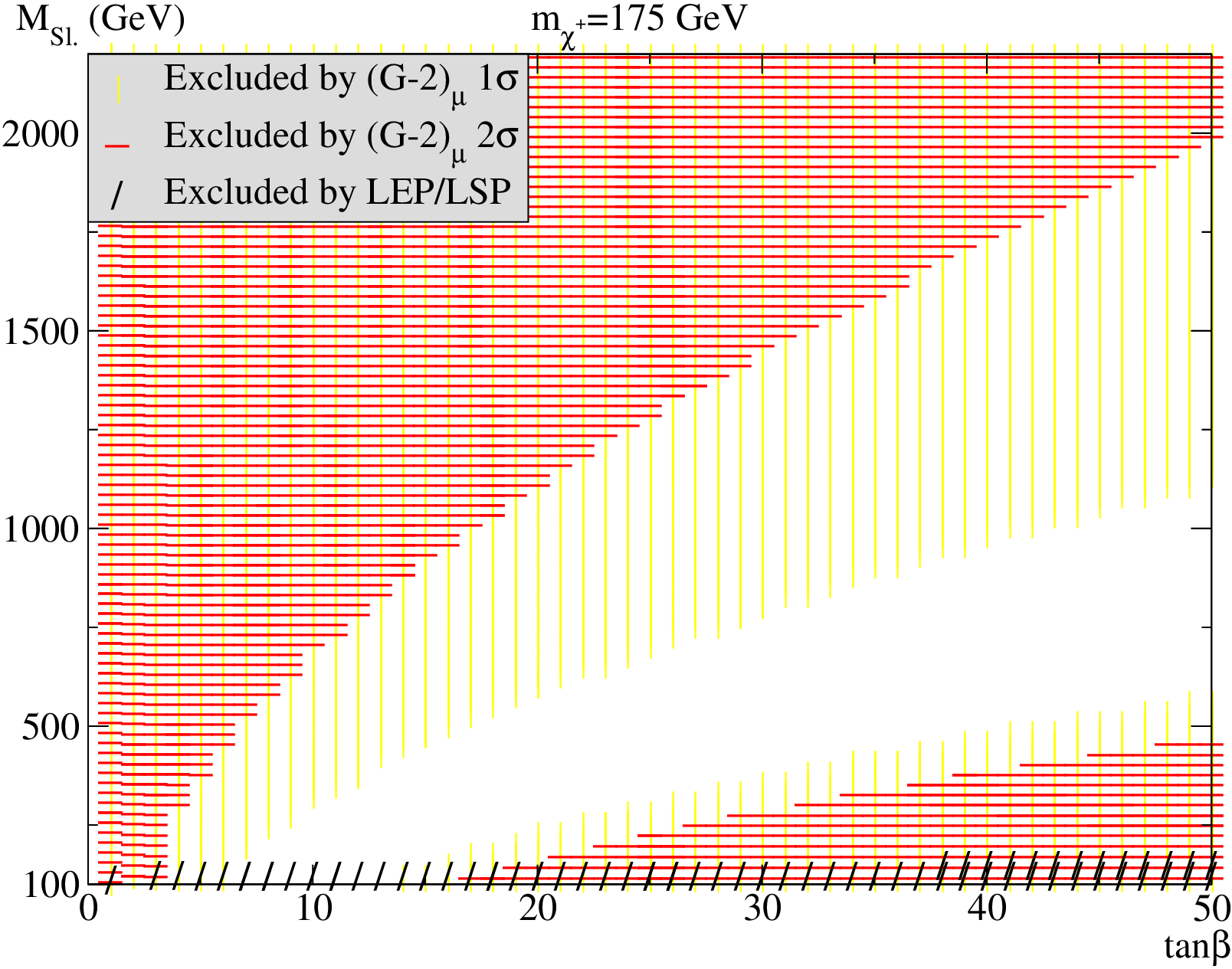}
\end{center}
\caption{Exclusion plot in the $\tan\beta$/slepton mass plane for a
light chargino of 175~GeV.}
\end{figure}

In Fig.~2 we present an exclusion plot in the $\tan\beta$/slepton mass
plane for a light chargino of 175~GeV. In the red and yellow hatched
domains the muon magnetic moment differs from the BNL result at the
$2\sigma$ and $1\sigma$ level, respectively. (In the excluded dark
hatched region a slepton would be the LSP, or violate constraints from
its non-observation at LEP.) Two different domains are excluded by
$a_{\mu}$: 

\begin{itemize}
\item In the bottom right-hand corner of Fig.~2, $\tan\beta$ is large
and sleptons are light. In this case, the chargino/sneutrino
contribution is strongly enhanced and becomes too large.
\item When sleptons are heavy and $\tan\beta$ is low (top and left-hand
side of Fig.~2), the chargino contribution to $a_{\mu}$ is suppressed,
and SUSY cannot explain the BNL result.
\end{itemize}

In order to see the impact of the chargino mass, we consider the case
of a heavier chargino of 450~GeV in Fig.~3. Here we chose 
$M_3=3M_2=6M_1=1.5$~TeV. One sees that light sleptons and large values
of $\tan\beta$ are required in order to account for the BNL result.

\begin{figure}[ht!]
\begin{center}
\includegraphics[width=13 cm]{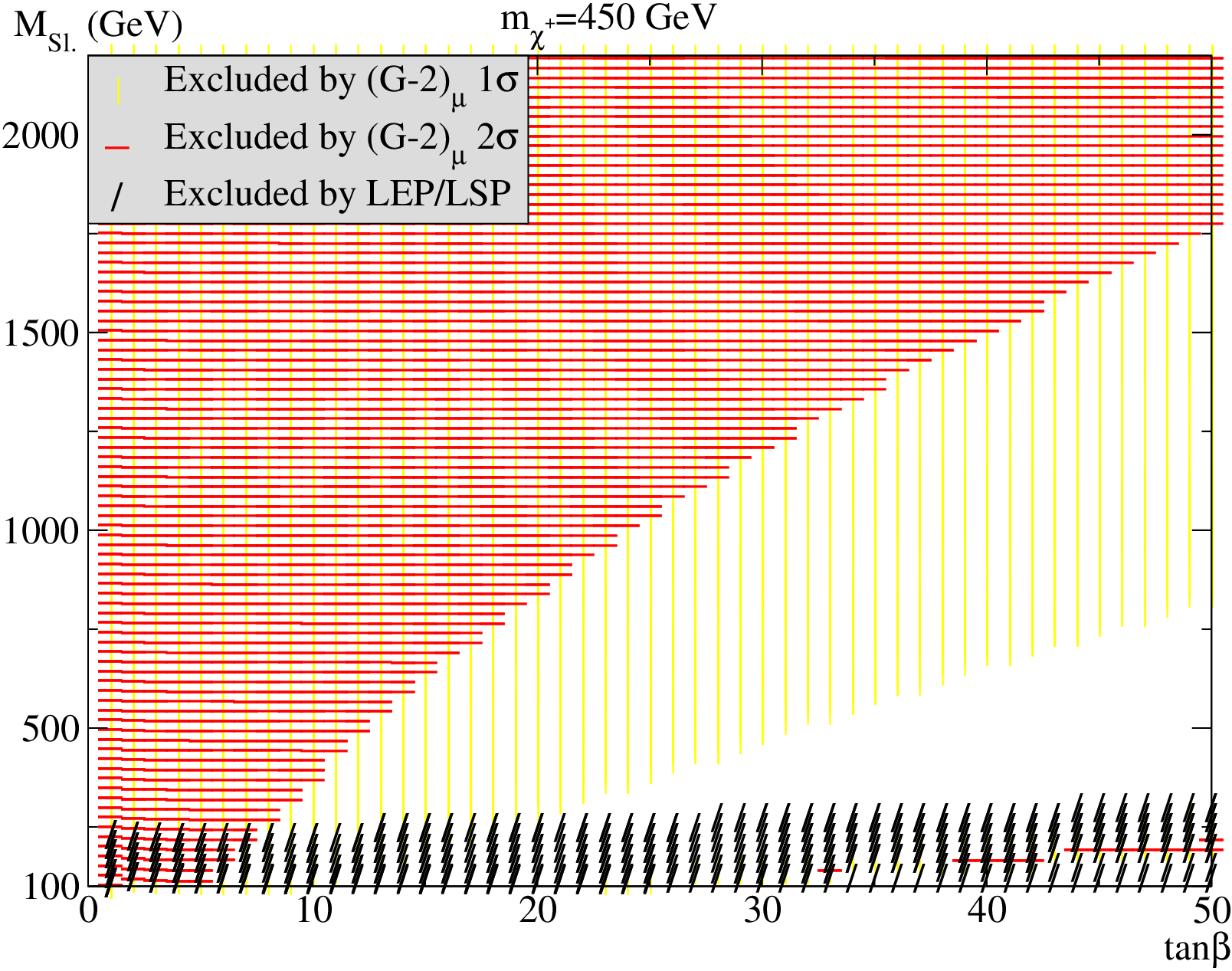}
\end{center}
\caption{Exclusion plot in the $\tan\beta$/slepton mass plane for a
heavier chargino of 450~GeV.}
\end{figure}

These results confirm that the chargino/sneutrino contribution can
explain the $3\sigma$ deviation between $a_{\mu}^{SM}$ and
$a_{\mu}^{EXP}$ in the NMSSM as well as in the MSSM, provided that the
supersymmetric particles are sufficiently light or $\tan\beta$ is
large.

Next we consider the NMSSM with a light pseudoscalar $a_1$ in the
spectrum. From the formulae (\ref{amuhiggs1L}) and (\ref{amuhiggs2L})
one finds that its contribution to $a_{\mu}$ depends essentially on the
pseudoscalar mass $m_{a_1}$ and its coupling to (down) fermions
$X_d=P'_{11}\tan\beta$, where $P'_{11}$ describes the mixing of $a_1$
with the Higgs-doublet sector. Subsequently we chose parameters in
the Higgs sector such that $P'_{11}$ remains approximately constant 
$\sim0.52$; then $X_d$ is entirely determined by $\tan\beta$.

In Fig.~4, we plot the contribution to $a_\mu$ which originates from
the NMSSM Higgs sector only against $\tan\beta$ (lower axis) or $X_d$
(upper axis) for $m_{a_1}=6.5$~GeV; then the contributions in section
2.2 are dominated by the light pseudoscalar. We indicate separately the
negative one-loop contribution (green curve), and the positive two-loop
contribution (red curve). For $m_{a_1}=6.5$~GeV the two-loop
contribution dominates, so that the total result (black curve) has the
same sign as the desired contribution (\ref{deviation}). The
contribution behaves as $\tan^2\beta$ (or $X_d^2$) and we find that the
Higgs contribution alone can reduce the discrepancy (\ref{deviation})
below the $2\sigma$ ($1\sigma$) level for  $X_d\gsim 15$ (22.5).

\begin{figure}[h!]
\begin{center}
\includegraphics[width=13 cm]{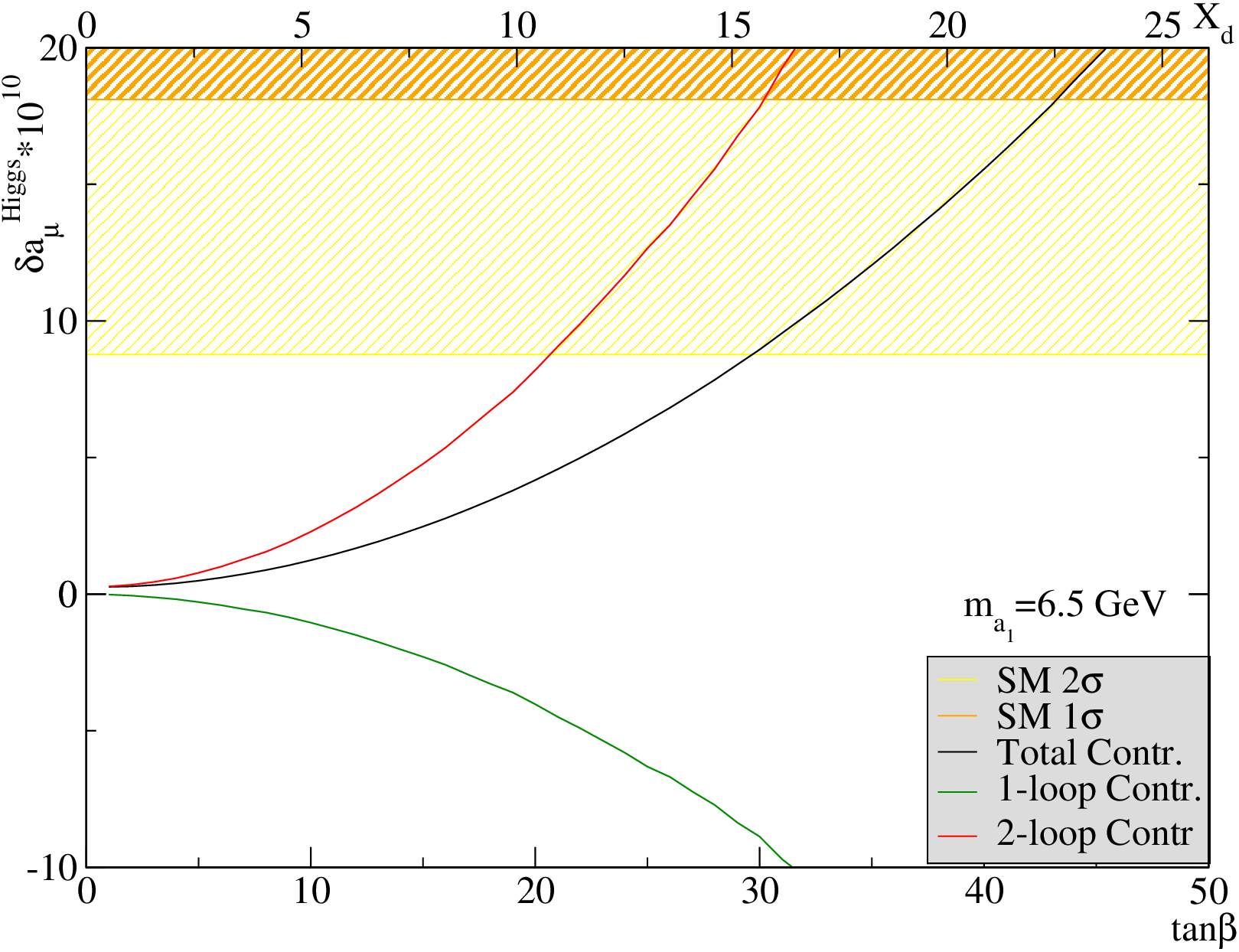}
\end{center}
\caption{Higgs contribution to $a_{\mu}$ as a function of $\tan\beta$
or $X_d$ for $m_{a_1} =$ 6.5~GeV.}
\end{figure}

In Fig.~5 we show the contribution to $a_\mu$ from the NMSSM Higgs
sector as a function of $m_{a_1}$ for various values of $X_d$. 
As we mentioned in the previous section, the negative one-loop
contribution dominates below $m_{a_1}\sim 3$~GeV, while the two-loop
diagram dominates for larger masses. Since both contributions decrease
when the pseudoscalar becomes heavy, the total Higgs contribution
becomes maximal around $\sim 6-7$~GeV.

Constraints from B physics (essentially from $\bar{B}_s\rightarrow
\mu^+\mu^-$ and $\Delta M_s$) depend mainly on the loop-induced
$b-s-a_1$ coupling \cite{dom}, and are particularly strong for $m_{a_1}
\sim M_{B_{d,s}} \sim 5$~GeV. For fixed $X_d$ (and sfermion masses and
trilinear couplings at $1$~TeV, $\lambda=0.3$, $\mu_{eff}=200$~GeV, and
$M_3=3M_2=6M_1=1.2$~TeV) we varied the remaining parameters 
$\tan\beta$, $\kappa$ and $A_\kappa$ in the Higgs sector of the NMSSM
and obtained regions near  $m_{a_1} \sim 5$~GeV, which are {\it always}
excluded -- these regions  are indicated as dashed parts of the curves
in Fig.~5. For larger $X_d$  and corresponding larger values of
$\tan\beta$, the forbidden regions are  larger. For $X_d=24$, the
forbidden region extends up to  $m_{a_1} \sim 6.5$~GeV. The region
$m_{a_1} \gsim 6.5$~GeV, where  the total Higgs contribution can be
relatively large, can be consistent with  B~physics constraints,
however.

\begin{figure}[h!]
\begin{center}
\includegraphics[width=13 cm]{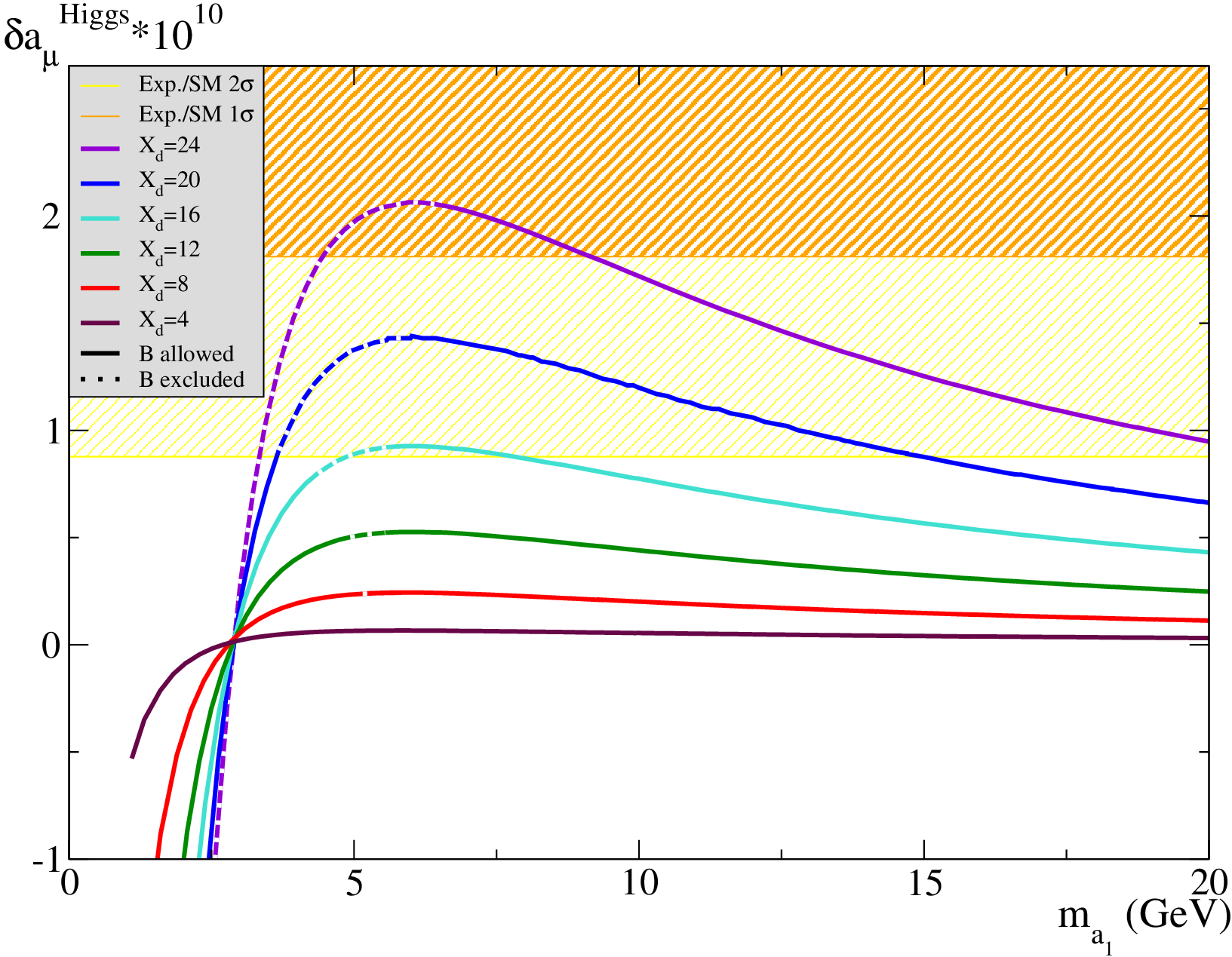}
\end{center}
\caption{Higgs contribution to $a_{\mu}$ as a function of the
pseudoscalar mass.}
\end{figure}

The previous results concern certain regions of the relatively large
parameter space of the general (low energy) NMSSM. Next we
consider the impact of $a_\mu$ on the cNMSSM with universal scalar and
gaugino masses ($m_0$ and $M_{1/2}$) as  well as universal trilinear
couplings ($A_0$) at the GUT~scale. Such a model is  motivated by
flavour-blind, supergravity-induced SUSY breaking scenarios.

A recent study of the cNMSSM parameter space spanned by $\lambda$,
$m_0$, $M_{1/2}$ and $A_0$ \cite{cNMSSM} showed that LEP constraints on
Higgs scalars and, notably, a dark matter relic density in agreement
with WMAP constraints require $\lambda \ll 1$, $m_0 \ll M_{1/2}$ and
that $A_0$ is determined in terms of $M_{1/2}$. Consequently, the
complete Higgs and sparticle spectrum depends essentially only on
$M_{1/2}$, which can {\it a priori} vary between 400~GeV and 2--3~TeV
(where all other constraints on sparticle masses and from B-physics are
satisfied; the lower limit on $M_{1/2}$ originates simultaneously from
the lower experimental bound on stau masses and LEP constraints on
Higgs scalars).

Using the most recent version of NMSSMTools \cite{nmssmtools}, we have
computed the spectrum and the supersymmetric contributions to $a_{\mu}$ 
in the cNMSSM as a function of $M_{1/2}$, with the result shown in
Fig.~6.

\begin{figure}[ht]
\begin{center}
\includegraphics[width=13 cm]{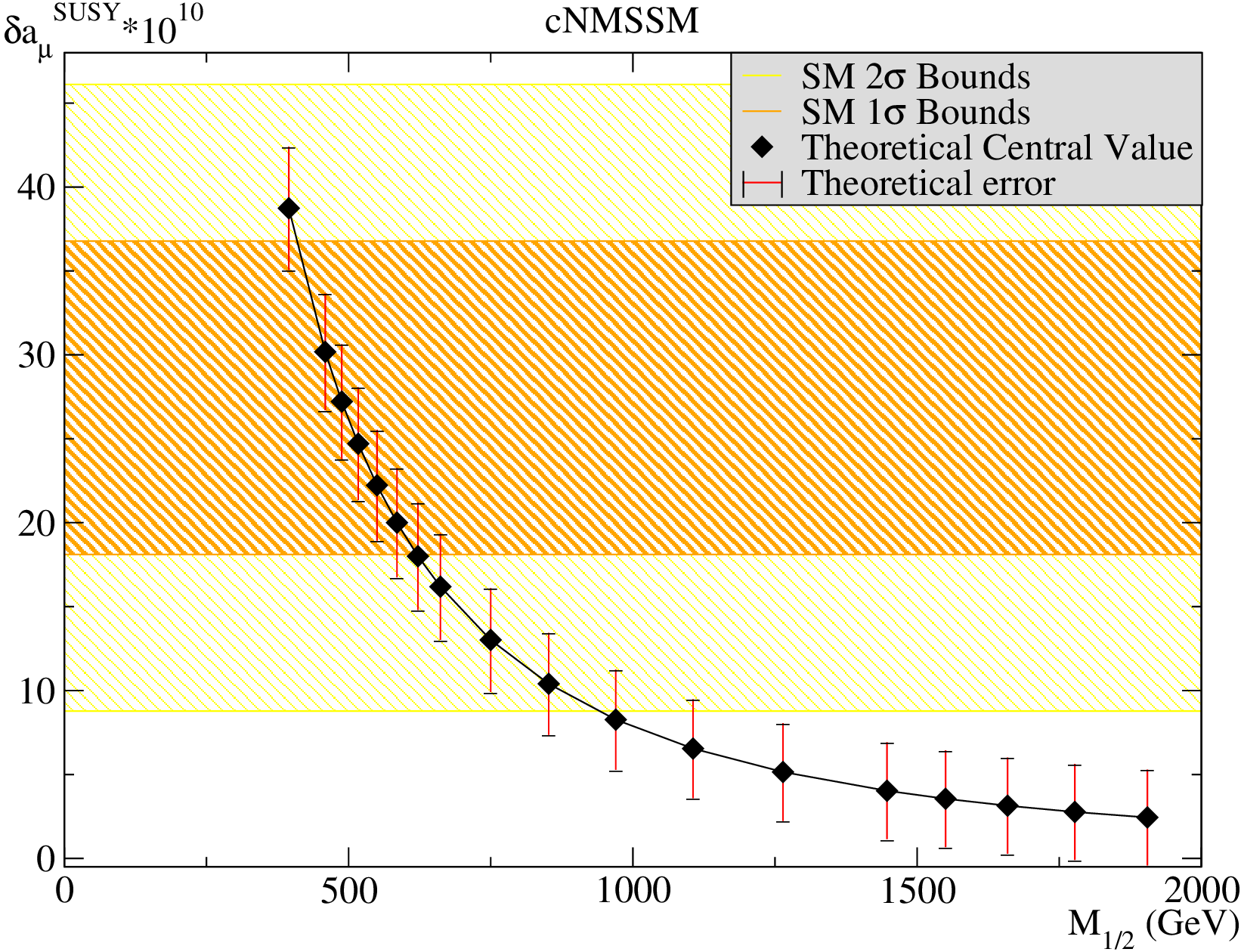}
\end{center}
\caption{$\delta a_{\mu}$ as a function of $M_{1/2}$ in the cNMSSM}
\end{figure}

We see that the constraint from $a_\mu$ confines the allowed range of
$M_{1/2}$ to $M_{1/2}\lsim 1$~TeV at the $2\sigma$ level, and to
400~GeV~$< M_{1/2} \lsim$ 700~GeV (where the sparticle spectrum is not
too heavy) at the $1\sigma$ level. In fact, the present experimental
value could be matched to arbitrarily high precision, and a more
precise measurement of $a_\mu$ could determine $M_{1/2}$ completely.

In the cNMSSM, the leading SUSY contributions to $a_\mu$ originate
from the MSSM-like one-loop diagrams. Besides the chargino/sneutrino
loop, the bino/smuon contribution is also quite significant. On the
other hand, contributions from the Higgs sector are always negligible in
the cNMSSM, since the lightest pseudoscalar is not very light.

To summarize, the deviation (\ref{deviation}) of the measured value of
$a_\mu$ from the SM can be explained in the NMSSM as in the MSSM since
-- in the absence of a light CP-odd Higgs scalar -- the corresponding
contributions are practically the same. In the presence of a light
CP-odd Higgs scalar the NMSSM specific contributions to $a_\mu$ are not
negligible in general (depending on $X_d$ and $m_{a_1}$, see Figs.~4
and 5) and, for $m_{a_1} >$ 3~GeV, can allow for a heavier sparticle
spectrum. The mSUGRA-like cNMSSM is consistent with constraints from
$a_\mu$, which can help to confine the remaining free parameter
$M_{1/2}$.

\newpage

%\appendix
\section*{ Appendix: Loop functions}
The functions appearing in the one-loop contributions (\ref{amuchar},
\ref{amuneutr}) are given in \cite{Martin}:
\begin{eqnarray*}
F^N_1(x) & = &\frac{2}{(1-x)^4}
\left[ 1-6x+3x^2+2x^3-6x^2\ln x\right] \\
F^N_2(x) & = &\frac{3}{(1-x)^3}\left[ 1-x^2+2x\ln x\right] \\
F^C_1(x) & = &\frac{2}{(1-x)^4}\left[ 2+ 3x - 6
x^2 + x^3 +6x\ln x\right] \\
F^C_2(x) & = & -\frac{3}{2(1-x)^3}\left[ 3-4x+x^2
    +2\ln x\right],
\end{eqnarray*}
The functions appearing in the two-loop contributions (\ref{Sfer},
\ref{char}, 2.23, \ref{amuhiggs2L}) are as in \cite{stock}:
\begin{eqnarray*}
f_{\tilde{f}}(z)&=&-\frac{z}{2}\int_0^1{\frac{x(1-x)}{x(1-x)-z}
\ln \frac{x(1-x)}{z}\,dx}=\frac{z}{2}\left[2+\ln z-f_{PS}(z)\right]\\
f_{PS}(z)&=&z\int_0^1{\frac{1}{x(1-x)-z}\ln \frac{x(1-x)}{z}\,dx}\\
&=&\frac{2z}{\sqrt{1-4z}}\Big[{\rm Li}_2
\Big(1-\frac{1-\sqrt{1-4z}}{2z}\Big)
-{\rm Li}_2\Big(1-\frac{1+\sqrt{1-4z}}{2z}\Big) \Big]\\
f_S(z)&=&-z\int_0^1{\frac{1-2x(1-x)}{x(1-x)-z}
\ln \frac{x(1-x)}{z}\,dx}\\
&=&(2z-1)f_{PS}(z)-2z\left(\ln z +2\right)
\end{eqnarray*}


\begin{thebibliography}{99}

\bibitem{bennett}
  G.~W.~Bennett {\it et al.}  [Muon G-2 Collaboration],
  Phys.\ Rev.\  D {\bf 73} (2006) 072003
  [arXiv:hep-ex/0602035].

\bibitem{jeger1}
  F.~Jegerlehner,
  Nucl.\ Phys.\ Proc.\ Suppl.\  {\bf 162} (2006) 22
  [arXiv:hep-ph/0608329].
 
\bibitem{hagiwara}
  K.~Hagiwara, A.~D.~Martin, D.~Nomura and T.~Teubner,
  Phys.\ Lett.\  B {\bf 649} (2007) 173
  [arXiv:hep-ph/0611102].

\bibitem{davier}
  M.~Davier,
  %``The hadronic contribution to (g-2)(mu),''
  Nucl. Phys. Proc. Suppl. {\bf 169} (2007) 288
  [arXiv:hep-ph/0701163].
 
\bibitem{jeger2}
  F.~Jegerlehner,
  %``Essentials of the Muon g-2,''
  Acta Phys.\ Polon.\  B {\bf 38} (2007) 3021
  [arXiv:hep-ph/0703125].

\bibitem{zhang}
  Z.~Zhang,
  ``Muon g-2: a mini review,''
  arXiv:0801.4905 [hep-ph].
  
\bibitem{Alemany}
  R.~Alemany, M.~Davier and A.~Hocker,
  Eur.\ Phys.\ J.\  C {\bf 2}, 123 (1998)
  [arXiv:hep-ph/9703220].

\bibitem{Daviertau}
  M.~Davier, S.~Eidelman, A.~Hocker and Z.~Zhang,
  Eur.\ Phys.\ J.\  C {\bf 31}, 503 (2003)
  [arXiv:hep-ph/0308213].

\bibitem{stock}
  D.~Stockinger,
  %``The muon magnetic moment and supersymmetry,''
  J.\ Phys.\ G {\bf 34} (2007) R45
  [arXiv:hep-ph/0609168].
  
\bibitem{1loop}
  P. Fayet, in {\it Unification of the Fundamental Particles
  Interactions}, eds. S.~Ferrara, J.~Ellis, P.~van Nieuwenhuizen,
  (Plenum, NY, 1980), p.~587;\\
  J.~A.~Grifols and A.~Mendez, Phys.\ Rev.\  D {\bf 26} (1982) 1809;\\
  J.~R.~Ellis, J.~S.~Hagelin and D.~V.~Nanopoulos,
  Phys.\ Lett.\  B {\bf 116} (1982) 283;\\
  R.~Barbieri and L.~Maiani, Phys.\ Lett.\  B {\bf 117} (1982) 203;\\
  D.~A.~Kosower, L.~M.~Krauss and N.~Sakai,
  Phys.\ Lett.\  B {\bf 133} (1983) 305;\\
  T.~C.~Yuan, R.~Arnowitt, A.~H.~Chamseddine and P.~Nath,
  Z.\ Phys.\  C {\bf 26} (1984) 407;\\
  J.~C.~Romao, A.~Barroso, M.~C.~Bento and G.~C.~Branco,
  Nucl.\ Phys.\  B {\bf 250} (1985) 295;\\
  I.~Vendramin, Nuovo Cim.\  A {\bf 101} (1989) 731;\\
  S.~A.~Abel, W.~N.~Cottingham and I.~B.~Whittingham,
  Phys.\ Lett.\  B {\bf 259} (1991) 307;\\
  U.~Chattopadhyay and P.~Nath,
  Phys.\ Rev.\  D {\bf 53} (1996) 1648 [arXiv:hep-ph/9507386];\\
  T.~Moroi, Phys.\ Rev.\  D {\bf 53} (1996) 6565 [Erratum-ibid.\  D
  {\bf 56} (1997) 4424] [arXiv:hep-ph/9512396].

\bibitem{Martin}
  S.~P.~Martin and J.~D.~Wells,
  Phys.\ Rev.\  D {\bf 64}, 035003 (2001)
  [arXiv:hep-ph/0103067].

\bibitem{byrne}
  M.~Byrne, C.~Kolda and J.~E.~Lennon,
  Phys.\ Rev.\  D {\bf 67} (2003) 075004
  [arXiv:hep-ph/0208067].

\bibitem{martinw}
  S.~P.~Martin and J.~D.~Wells,
  Phys.\ Rev.\  D {\bf 67} (2003) 015002
  [arXiv:hep-ph/0209309].

 \bibitem{Degrassi}
  G.~Degrassi and G.~F.~Giudice,
  Phys.\ Rev.\  D {\bf 58}, 053007 (1998)
  [arXiv:hep-ph/9803384].

\bibitem{chang} 
D.~Chang, W.~F.~Chang, C.~H.~Chou and W.~Y.~Keung,
  Phys.\ Rev.\  D {\bf 63} (2001) 091301
  [arXiv:hep-ph/0009292].
 
\bibitem{Cheung}
  K.~M.~Cheung, C.~H.~Chou and O.~C.~W.~Kong,
  Phys.\ Rev.\  D {\bf 64}, 111301 (2001)
  [arXiv:hep-ph/0103183].

\bibitem{Chen}
  C.~H.~Chen and C.~Q.~Geng, Phys.\ Lett.\  B {\bf 511} (2001) 77
  [arXiv:hep-ph/0104151].
  
\bibitem{Arhrib}
  A.~Arhrib and S.~Baek,
  Phys.\ Rev.\  D {\bf 65}, 075002 (2002)
  [arXiv:hep-ph/0104225].

\bibitem{HSW1}
  S.~Heinemeyer, D.~Stockinger and G.~Weiglein,
  Nucl.\ Phys.\  B {\bf 690}, 62 (2004)
  [arXiv:hep-ph/0312264].

\bibitem{HSW2}
  S.~Heinemeyer, D.~Stockinger and G.~Weiglein,
  Nucl.\ Phys.\  B {\bf 699}, 103 (2004)
  [arXiv:hep-ph/0405255].
  
\bibitem{fengli}
    T.~F.~Feng, X.~Q.~Li, L.~Lin, J.~Maalampi and H.~S.~Song,
  Phys.\ Rev.\  D {\bf 73} (2006) 116001 [arXiv:hep-ph/0604171].
  
\bibitem{Fengsun}
  T.~F.~Feng, L.~Sun and X.~Y.~Yang, ``Electroweak and supersymmetric
  two-loop corrections to lepton anomalous magnetic and electric dipole
  moments,'' arXiv:0805.1122 [hep-ph].

\bibitem{ghm}
  J.~F.~Gunion, D.~Hooper and B.~McElrath,
  %``Light neutralino dark matter in the NMSSM,''
  Phys.\ Rev.\  D {\bf 73}, 015011 (2006)
  [arXiv:hep-ph/0509024].
  
\bibitem{bklh}
V.~Barger, C.~Kao, P.~Langacker and H.~S.~Lee, Phys.\ Lett.\  B {\bf
  614} (2005) 67 [arXiv:hep-ph/0412136].

\bibitem{nmssm} H.P. Nilles, M. Srednicki and D. Wyler, Phys. Lett. B
\textbf{120} (1983) 346;\\ 
J.M. Frere, D.R. Jones and S. Raby, Nucl. Phys. B \textbf{222} (1983)
11;\\ 
J.P. Derendinger and C.A. Savoy, Nucl. Phys. B \textbf{237} (1984)
307;\\  
J.R. Ellis, J.F. Gunion, H.E. Haber, L. Roszkowski and F. Zwirner,
Phys. Rev. D \textbf{39} (1989) 844;\\
M. Drees, Int. J. Mod. Phys. A \textbf{4} (1989) 3635.

\bibitem{light1}
B.~A.~Dobrescu, G.~Landsberg and K.~T.~Matchev, Phys.\ Rev.\ D {\bf 63}
  (2001) 075003 [arXiv:hep-ph/0005308];\\
B.~A.~Dobrescu and K.~T.~Matchev, JHEP {\bf 0009} (2000) 031 
  [arXiv:hep-ph/0008192].

\bibitem{light2}
U.~Ellwanger, J.~F.~Gunion, C.~Hugonie and S.~Moretti,
arXiv:hep-ph/0305109 (in  ``Physics interplay of the LHC and the ILC'',
G.~Weiglein {\it et al.}  [LHC/LC Study Group], Phys.\ Rept.\  {\bf
426} (2006) 47); arXiv:hep-ph/0401228 (in ``The Higgs working group:
Summary report 2003'', K.~A.~Assamagan {\it et al.}  [Higgs Working
Group Collaboration],  arXiv:hep-ph/0406152);\\  
R.~Dermisek and J.~F.~Gunion,
  Phys.\ Rev.\ Lett.\  {\bf 95} (2005) 041801 [arXiv:hep-ph/0502105],
  Phys.\ Rev.\ D {\bf 73} (2006) 111701 
  [arXiv:hep-ph/0510322], Phys.\ Rev.\  D {\bf 75} (2007) 075019
  [arXiv:hep-ph/0611142];\\
U.~Ellwanger, J.~F.~Gunion and C.~Hugonie, JHEP {\bf 0507} (2005) 041 
 [arXiv:hep-ph/0503203];\\
S.~Chang, P.~J.~Fox and N.~Weiner,
  JHEP {\bf 0608} (2006) 068 [arXiv:hep-ph/0511250];\\
P.~W.~Graham, A.~Pierce and J.~G.~Wacker, ``Four taus at the
  Tevatron,'' arXiv:hep-ph/0605162;\\
S.~Moretti, S.~Munir and P.~Poulose, Phys.\ Lett.\ B {\bf 644} (2007)
  241  [arXiv:hep-ph/0608233];\\
S.~Chang, P.~J.~Fox and N.~Weiner,
  Phys.\ Rev.\ Lett.\  {\bf 98} (2007) 111802 [arXiv:hep-ph/0608310];\\
T.~Stelzer, S.~Wiesenfeldt and S.~Willenbrock,
  Phys.\ Rev.\  D {\bf 75} (2007) 077701 [arXiv:hep-ph/0611242];\\
U.~Aglietti {\it et al.}, ``Tevatron-for-LHC report: Higgs,''
  arXiv:hep-ph/0612172;\\
E.~Fullana and M.~A.~Sanchis-Lozano,
  Phys.\ Lett.\  B {\bf 653} (2007) 67
  [arXiv:hep-ph/0702190];\\  
K.~Cheung, J.~Song and Q.~S.~Yan,
  Phys.\ Rev.\ Lett.\  {\bf 99} (2007) 031801 [arXiv:hep-ph/0703149];\\
M.~A.~Sanchis-Lozano,
  ``A light non-standard Higgs boson: to be or not to be at a (Super) B
  factory?,'' arXiv:0709.3647 [hep-ph];\\
M.~Carena, T.~Han, G.~Y.~Huang and C.~E.~M.~Wagner, JHEP {\bf 0804}
  (2008) 092 [arXiv:0712.2466 [hep-ph]];\\
J.R. Forshaw, J.F. Gunion, L. Hodgkinson, A. Papaefstathiou and  A.D.
   Pilkington, arXiv:0712.3510 [hep-ph];\\
Z.~Heng, R.~J.~Oakes, W.~Wang, Z.~Xiong and J.~M.~Yang, ``B meson
  Dileptonic Decays in NMSSM with a Light CP-odd Higgs Boson,''
  arXiv:0801.1169 [hep-ph];\\
A.~Djouadi {\it et al.}, ``Benchmark scenarios for the NMSSM'',
  arXiv:0801.4321 [hep-ph];\\
X.~G.~He, J.~Tandean and G.~Valencia, ``Rare Decays with a Light CP-Odd
  Higgs Boson in the NMSSM,'' arXiv:0803.4330 [hep-ph].


\bibitem{kraw}
  M.~Krawczyk,
  Acta Phys.\ Polon.\  B {\bf 33}, 2621 (2002)
  [arXiv:hep-ph/0208076].

\bibitem{lep}
 S.~Schael {\it et al.}  [ALEPH, DELPHI, L3 and OPAL Collaborations],
  Eur.\ Phys.\ J.\  C {\bf 47} (2006) 547

\bibitem{dom}
F.~Domingo and U.~Ellwanger,
  %``Updated Constraints from B Physics on the MSSM and the NMSSM,''
  JHEP {\bf 0712} (2007) 090
  [arXiv:0710.3714 [hep-ph]].
 
\bibitem{4lqed}
  S.~Laporta and E.~Remiddi,
  Phys.\ Lett.\  B {\bf 379}, 283 (1996)
  [arXiv:hep-ph/9602417],
  Phys.\ Lett.\  B {\bf 301}, 440 (1993);\\
  T.~Aoyama, M.~Hayakawa, T.~Kinoshita and M.~Nio,
  %``Revised value of the eighth-order electron g-2,''
  Phys.\ Rev.\ Lett.\  {\bf 99}, 110406 (2007)
  [arXiv:0706.3496 [hep-ph]];\\
  T.~Kinoshita and M.~Nio,
  %``Improved alpha**4 term of the muon anomalous magnetic moment,''
  Phys.\ Rev.\  D {\bf 70}, 113001 (2004)
  [arXiv:hep-ph/0402206].

\bibitem{5lqed}
  T.~Kinoshita and M.~Nio,
  Phys.\ Rev.\  D {\bf 73}, 053007 (2006)
  [arXiv:hep-ph/0512330].

\bibitem{bijnens}
  J.~Bijnens and J.~Prades,
  Mod.\ Phys.\ Lett.\  A {\bf 22} (2007) 767
  [arXiv:hep-ph/0702170]. 
 
\bibitem{czarn}
  A.~Czarnecki, W.~J.~Marciano and A.~Vainshtein,
  Phys.\ Rev.\  D {\bf 67} (2003) 073006
  [Erratum-ibid.\  D {\bf 73} (2006) 119901]
  [arXiv:hep-ph/0212229].

\bibitem{nmssmtools}
  U.~Ellwanger, J.~F.~Gunion and C.~Hugonie,
  JHEP {\bf 0502} (2005) 066
  [arXiv:hep-ph/0406215];\\
  U.~Ellwanger and C.~Hugonie,
  Comput.\ Phys.\ Commun.\  {\bf 177} (2007) 399,\\
  {\sf http://www.th.u-psud.fr/NMHDECAY/nmssmtools.html}

\bibitem{slha}
  P.~Skands {\it et al.},
  JHEP {\bf 0407} (2004) 036
  [arXiv:hep-ph/0311123].

\bibitem{slha2}
  B.~Allanach {\it et al.},
  ``SUSY Les Houches Accord 2,''
  arXiv:0801.0045 [hep-ph].
  
\bibitem{Leveille}
  J.~P.~Leveille, Nucl.\ Phys.\  B {\bf 137} (1978) 63.

\bibitem{cNMSSM}
  A.~Djouadi, U.~Ellwanger and A.~M.~Teixeira,
  ``The constrained next-to-minimal supersymmetric standard model,''
  arXiv:0803.0253 [hep-ph].

\end{thebibliography}
\end{document}